\documentstyle[12pt]{article}

\begin{document}

\title{Planar Dirac Electron
in Coulomb and Magnetic Fields:\\ a Bethe ansatz approach}

\author{Chun-Ming Chiang$^{1,2}$ and Choon-Lin Ho$^1$}
\date
{\small \sl $^1$Department of Physics, Tamkang University, Tamsui
25137, Taiwan\\ $^2$Kuang Wu Institute of Technology, Peitou,
Taipei 112, Taiwan}

\maketitle

\begin{abstract}
The Dirac equation for an electron in two spatial dimensions in
the Coulomb and homogeneous magnetic fields is an example of the
so-called quasi-exactly solvable models. The solvable parts of its
spectrum was previously solved from the recursion relations.  In
this work we present a purely algebraic solution based on the
Bethe ansatz equations.  It is realised that, unlike the
corresponding problems in the Schr\"odinger and the Klein-Gordon
case, here the unknown parameters to be solved for in the Bethe
ansatz equations include not only the roots of wave function
assumed, but also a parameter from the relevant operator. We also
show that the quasi-exactly solvable differential equation does
not belong to the classes based on the algebra $sl_2$.
\end{abstract}

\vskip0.3cm
\noindent{May 22, 2001}
\newpage

{\bf 1.}  Recently a new type of spectral problem, so-called
quasi-exactly solvable model (QESM), was discovered by physicists
and mathematicians [\cite{TU}-\cite{Zas}]. This is a special class
of quantum-mechanical problems for which analytical solutions are
possible only for parts of the energy spectra and for particular
values of the fundamental parameters. The reason for such
quasi-exactly solvability is usually the existence of a hidden
Lie-algebraic structure [\cite{Tur2}-\cite{KO}].  More precisely,
quasi-exactly solvable (QES) Hamiltonian can be reduced to a
quadratic combination of the generators of a Lie group with
finite-dimensional representations.

The first physical example of QESM in atomic physics is the system
of two electrons moving in an external oscillator potential
discussed in \cite{SG,Taut1}.  The authors of these works
apparently were unaware of the mathematical development in QESM.
Later, several physical QESM were discovered, which include the
two-dimensional Schr\"odinger \cite{Taut2}, the Klein-Gordon
\cite{VP}, and the Dirac equations \cite{HoKha} of an electron
moving in an attractive/repulsive Coulomb field and a homogeneous
magnetic field. The essential features shared by all these above
examples are as follows. The differential equations are solved
according to the standard procedure.  After separating out the
asymptotic behaviors of the system,  one obtains an equation for
the part which can be expanded as a power series of the basic
variable. But instead of the two-step recursion relations for the
coefficients of power series so often encountered in exactly
solvable problems, one gets three-step recursion relations.  The
complexity of the recursion relations does not allow one to
determine the energy spectrum exactly from the normalisability of
the eigenfunctions. However, one can impose a sufficient condition
for normalisability by terminating the series at a certain order
of power of the variable; {\it i.e.} by choosing a polynomial. By
doing so one could obtain exact solutions to the original problem,
but only for certain energies and for specific values of the
parameters of the problem. These parameters, namely, are the
frequency of the oscillator potential and the external magnetic
fields.

In \cite{ChHo} a systematic and unified algebraic treatment was
given to the above-mentioned systems, with the exception of the
Dirac case.  This was made possible by realising that these
systems are governed essentially by the same basic equation, which
is quasi-exactly solvable owing to the existence of a hidden
$sl_2$ algebraic structure.  This algebraic structure was first
realised by Turbiner for the case of two electrons in an
oscillator potential \cite{Tur1}.  In this algebraic approach,
analytic expressions of the solvable parts of the energy spectrum
and the allowed parameters were expressible in terms of the roots
of a set of Bethe ansatz equations.

In this paper we would like to extend the method of \cite{ChHo} to
the planar Dirac equation of an electron in the Coulomb and
magnetic fields. It turns out the a set of Bethe ansatz equation
can also be set up in this case. However, unlike the systems
considered in \cite{ChHo}, here the unknown variables in the Bethe
ansatz equations involved not only the roots of the wave functions
assumed, but also a parameter from the relevant operator.  We also
demonstrate that the Bethe ansatz approach yields the same
spectrum as that obtained by solving recursion relations. Finally,
we show that the quasi-exactly solvability of this system is not
related to the $sl_2$ algebra.

\vskip 0.5 truecm

{\bf 2.}  In 2+1 dimensions
the Dirac algebra
\begin{eqnarray}
 \{\gamma^{\mu}, \gamma^{\nu}\} = 2g^{\mu\nu}~,~~~~g^{\mu\nu}= {\rm
diag}(1,-1,-1)
\label{anti}
\end{eqnarray}
may be represented in terms of
the Pauli matrices as $\gamma^0 = \sigma_3$, $\gamma^k = i\sigma_k$, or
equivalently, the matrices $(\alpha_1,\alpha_2)=\gamma^0
(\gamma^1,\gamma^2)=(-\sigma_2,\sigma_1)$ and $\beta=\gamma^0$.
Then the Dirac  equation for an electron minimally coupled to an external
electromagnetic field has the form (we set $c=\hbar=1$)
\begin{eqnarray}
(i\partial_t - H_D)\Psi(t, {\bf r}) = 0,
\label{eq2}
\end{eqnarray}
where
\begin{eqnarray}
  H_D = {\bf \alpha}{\bf P} + \beta m - eA^0 \equiv \sigma_1P_2 -
\sigma_2P_1 + \sigma_3m - eA^0
\label{eq3}
\end{eqnarray}
is the Dirac Hamiltonian, $P_{k} = -i\partial_k + eA_k$ is the
operator of generalized momentum of the electron, $A_{\mu}$ the
vector potential of the external electromagnetic field, $m$ the
rest mass of the electron, and $-e~ (e>0)$ is its electric charge.
The Dirac wave function $\Psi(t, {\bf r})$ is a two-component
function. In an external Coulomb field and a constant homogeneous
magnetic field $B>0$ along the $z$ direction, the potential
$A_\mu$ assume the following forms in the symmetric gauge
\begin{eqnarray}
 A^0(r) = Ze/r~ (e>0), \quad A_x = -By/2, \quad A_y = Bx/2~.
\label{e1}
\end{eqnarray}
We assume the wave functions to have the form
\begin{eqnarray}
 \Psi(t,{\bf x}) = \frac{1}{\sqrt{r}}\exp(-iEt)
\psi_l(r, \varphi)~,
\label{e3}
\end{eqnarray}
where  $E$ is the energy of the electron, and
\begin{eqnarray}
\psi_l(r, \varphi) =
\left( \begin{array}{c}
F(r)e^{il\varphi}\\
G(r)e^{i(l+1)\varphi}
\end{array}\right)
\label{eqn6}
\end{eqnarray}
with integral number $l$.
The function $\psi_l(r,\varphi)$ is an eigenfunction of the conserved total
angular momentum $J_z=L_z + S_z = -i\partial/\partial\varphi + \sigma_3/2$
with eigenvalue $j=l+1/2$.    It should be reminded that $l$ is not a
good quantum number.  Only the eigenvalues $j$ of the conserved total angular
momentum $J_z$ are physically meaningful.

By putting Eq.(\ref{e3}) and (\ref{eqn6}) into (\ref{eq2}), and
taking into account of the equations
\begin{eqnarray}
 P_x \pm iP_y = -ie^{\pm i\varphi}\left(\frac{\partial}{\partial r} \pm
\left(\frac{i}{r}\frac{\partial}{\partial\varphi}-\frac{eBr}{2}\right)
\right)~,
\label{impul}
\end{eqnarray}
we obtain
\begin{eqnarray}
\frac{dF}{dr} - \left(\frac{l+\frac{1}{2}}{r} + \frac{eBr}{2}\right)F +
\left(E + m + \frac{Z\alpha}{r}\right)G = 0~,
\label{d1} \\
\frac{dG}{dr} + \left(\frac{l+\frac{1}{2}}{r} + \frac{eBr}{2}\right)G - \left(
E - m + \frac{Z\alpha}{r}\right)F = 0~,
\label{d2}
\end{eqnarray}
where $\alpha\equiv e^2=1/137$ is the fine structure constant. In
a strong magnetic field the asymptotic solutions of $F(r)$ and
$G(r)$ have the forms $\exp(-eBr^2/4)$ at large $r$, and
$r^\gamma$ with $\gamma = \sqrt{(l+1/2)^2 - (Z\alpha)^2}$ for
small $r$. One must have $Z\alpha <1/2$, otherwise the wave
function will oscillate as $r\to 0$ when $l=0$ and $l=-1$.

Let us assume
\begin{eqnarray}
F(r)=r^\gamma\exp(-eBr^2/4)~Q(r),~~~~
G(r)=r^\gamma\exp(-eBr^2/4)~P(r)~.
\label{d3}
\end{eqnarray}
In \cite{HoKha} we showed that parts of the spectrum could be
analytically solved for by imposing the sufficient condition that
$Q(r)$ and $P(r)$ be polynomials, thus showing that the system
belongs to the QESM.  The spectrum was solved in \cite{HoKha} from
the recursion relations for the coefficients in the series
expansion in $Q$ and $P$.  In this paper, we will show that the
same spectrum can also be obtained in a purely algebraic way. This
is achieved by the method of factorisation which leads to a set of
Bethe ansatz equations \cite{HoKha,ChHo}.

Substituting  Eq.(\ref{d3}) into Eq.(\ref{d1}) and (\ref{d2}) and
eliminating $P(r)$ from the coupled equations, we have
\begin{eqnarray}
\left\{\frac{d^2}{dr^2}+\left[\frac{2\gamma}{r}-eBr+\frac{Z\alpha/r^2}
{E+m+Z\alpha/r}\right]\frac{d}{dr}
+E^2-m^2\right.~~~~~~~~\nonumber\\
+\frac{2EZ\alpha}{r}+\frac{l+\frac{1}{2}}{r^2}-\frac{\gamma}{r^2}-eB(\Gamma+
1 ) ~~~~~~~~~~~~~~\nonumber\\ \left.
+\frac{Z\alpha/r^2}{E+m+Z\alpha/r}\left[\frac{\gamma}{r}-eBr-\frac{l+1/
2 } {r}\right]\right\}~Q(r)=0~, \label{d4}
\end{eqnarray}
where $\Gamma = l + 1/2 + \gamma$.
Once $Q(r)$ is solved, the form of $P(r)$ is obtainable from Eqs.(\ref{d1})
and (\ref{d3}).
If we let $x=r/l_B$, $l_B=1/\sqrt{eB}$, Eq.(\ref{d4}) becomes
\begin{eqnarray}
\left\{\frac{d^2}{dx^2}+\left[\frac{2\gamma}{x}-x+\frac{Z\alpha}
{x((E+m)l_Bx+Z\alpha)}\right]\frac{d}{dx}\right.~~~~~~~~~~~~~~~\nonumber\\
+(E^2-m^2)l_B^2+\frac{2Ezl_B\alpha}{x}+\frac{(l+1/2-\gamma)}{x^2}
-(\Gamma+1)~~~~\nonumber\\
\left.-\frac{Z\alpha(l+1/2-\gamma)}{x^2\left[(E+m)l_Bx+Z\alpha\right]}
-\frac{Z\alpha}{(E+m)l_Bx+Z\alpha}\right\}~Q(x)=0~.
\label{d5}
\end{eqnarray}
Eq.(\ref{d5}) can be rewritten as
\begin{eqnarray}
\left\{\frac{d^2}{dx^2}+\left[\frac{2\beta}{x}-x-\frac{1}{x+x_0}\right]
\frac{d}{dx}+\epsilon+\frac{b}{x}-\frac{c}{x+x_0}\right\}~Q(x)=0~.
\label{d6}
\end{eqnarray}
Here $\beta=\gamma+1/2$, $x_0=Z\alpha/[(E+m)l_B]$,
$\epsilon=(E^2-m^2) l_B^2-(\Gamma+1)$, $b=b_0+L/x_0$,
$b_0=2EZ\alpha l_B$, $L=(l+1/2-\gamma)$,
 and $c=x_0+L/x_0$.  On expressing $l_B$ in the expression of $\epsilon$ in
terms of $x_0$, we get
\begin{eqnarray}
\epsilon=\frac{E-m}{E+m}\left(\frac{Z\alpha}{x_0}\right)^2
-\left(\Gamma+1\right)~.
\label{epsilon}
\end{eqnarray}
It is obvious that the energy $E$ is determined once we know the values of
$\epsilon$ and $x_0$.  The corresponding value of the magnetic field $B$ is
then obtainable from the expression $l_B=Z\alpha/[(E+m)x_0]$.  Solution of
$x_0$ is achieved below by means of the Bethe ansatz equations.

\vskip 0.5 truecm

{\bf 3.}  We observe that the problem of finding the spectrum for
Eq.(\ref{d6}) is equivalent to determining the eigenvalues of the
operator
\begin{eqnarray}
H=-\frac{d^2}{dx^2}-\left(\frac{2\beta}{x}-x-\frac{1}{x+x_0}\right)
\frac{d}{dx}-\frac{b}{x}+\frac{c}{x+x_0}~.
\label{d7}
\end{eqnarray}
We want to factorise the operator (\ref{d7}) in the form
\begin{eqnarray}
H=a^+a+\epsilon~.
\label{d8}
\end{eqnarray}
The eigenfunctions of the operator $H$ at $\epsilon=0$ must satisfy
the equation
\begin{eqnarray}
aQ(x)=0~.
\label{d9}
\end{eqnarray}
Suppose polynomial solution exist for Eq.(\ref{d6}), say $Q$
equals a non-vanishing constant, or $Q=\prod^n_{k=1}(x-x_k)$,
where $x_k$ are the zeros of $Q$, and $n$ is the degree of $Q$. In
the case where $Q$ is a constant (which may be viewed as
corresponding to $n=0$), the operators $a$ and $a^+$ have the form
\begin{eqnarray}
a=\frac{d}{d x}~,\quad\quad  a^+=-\frac{d}{d
x}-\left(\frac{2\beta}{x}-x-\frac{1}{x+x_0}\right)~.
\label{d10}
\end{eqnarray}
If $Q=\prod^n_{k=1}(x-x_k)$, $a$ and $a^+$ will assume the form
\begin{eqnarray}
a=\frac{d}{d x}-\sum^n_{k=1}\frac{1}{x-x_k}
\label{d12}
\end{eqnarray}
and
\begin{eqnarray}
a^+=-\frac{d}{d x}-\left(\frac{2\beta}{x}-x-\frac{1}{x+x_0}
\right)-\sum^n_{k=1}\frac{1}{x-x_k}~.
\label{d13}
\end{eqnarray}

We now substitute the forms of $a$ and $a^+$ into Eq.(\ref{d8})
and compare the result with Eq.(\ref{d7}).  This leads to
conditions that must be satisfied by the various parameters and
the roots $x_k$'s. For constant $Q$ ($n=0$), one has
\begin{eqnarray}
\epsilon=b=c=0~.
\label{BA1}
\end{eqnarray}
The fact that $c=0$ implies
\begin{eqnarray}
x_0^2=-L~.
\label{x0}
\end{eqnarray}
For $n\geq 1$, one gets
\begin{eqnarray}
b_0+\frac{L}{x_0}&=&2\beta\sum^n_{k=1}\frac{1}{x_k}~,\quad \quad
\epsilon=n~, \label{d18}\\
x_0+\frac{L}{x_0}&=&\sum^n_{k=1}\frac{1}{x_k+x_0}~, \label{d19}\\
\frac{2\beta}{x_k}-x_k &-& \frac{1}{x_k+x_0}-2\sum^n_{j\neq k}
\frac{1}{x_j-x_k}=0,\quad k=1,\dots,n~. \label{d20}
\end{eqnarray}
Eqs.(\ref{x0}), (\ref{d19}) and (\ref{d20}) constitute the set of
$n+1$ Bethe ansatz equations relevant to this Dirac system, which
involve $n+1$ unknown parameters $\{x_0,x_1,\ldots,x_n\}$ . It is
worthwhile to note that, unlike the corresponding equations in the
Schr\"odinger and the Klein-Gordon case discussed in \cite{ChHo},
this set of Bethe ansatz equations involved not only the roots
$x_k$'s, but also a parameter $x_0$ from $H$.  From the second
equation in Eq.(\ref{d18}) we get
\begin{eqnarray}
E^2 - m^2=\frac{1}{l_B^2}\left(\Gamma+n+1\right)~.
\label{E2}
\end{eqnarray}
Since $-1/2\leq \Gamma\leq 0$ for $Z\alpha<1/2$ \cite{HoKha}, we
see from Eq.(\ref{E2}) that the solvable parts of the spectrum
must satisfy $|E|\geq m$.

So we see that the solution of the solvable parts of the spectrum
$E$ boils down to solving the Bethe ansatz equations for $x_0$ in
the differential operator, and the roots $x_k$ ($k=1,\ldots,n$) of
$Q(x)$.  Once the value of $x_0$ for each order $n=\epsilon$ is
known, the energy $E$ is given by Eq.(\ref{epsilon}).  The
corresponding magnetic field $B$ is then determined from the
definition of $b_0$, or from Eq.(\ref{E2}).  The Bethe ansatz
equations thus provides a systematic solutions of the QES
spectrum.  Of course, as the order of the degree of $Q$ increases,
analytical solutions of the Bethe ansatz equations becomes
difficult, and one must resort to numerical methods.

\vskip 0.5 truecm

{\bf 4.}  In what follows we
shall show the consistency of the solutions
by the Bethe ansatz approach and that by the recursion relations
presented in \cite{HoKha} for the first three lowest orders
($n=0,1,2$) in $Q$. Instead of solving for $x_0$, our strategy is
to eliminate it in Eq.(\ref{epsilon}) by means of the equations
(\ref{x0})-(\ref{d20}) so as to obtain an equation obeyed by $E$
for each order of $Q$.  This equation is then compared with the
corresponding equation obtained from the recursion relations as
presented in \cite{HoKha}.

From Eq.(\ref{BA1}) and (\ref{x0}) we have $x_0^2=-L$ and
$\epsilon=0$ when $Q$ is a constant. Substitute these values of
$x_0$ and $\epsilon$ into Eq.(\ref{epsilon}), and using the fact
that $\Gamma L=(Z\alpha)^2$, we obtain the corresponding value of
$E$ as
\begin{eqnarray}
E=-\frac{m}{2(l+\gamma+1)}~.
\label{d16}
\end{eqnarray}
This is the result presented in \cite{HoKha}.  The corresponding
allowed value of the magnetic field $B$ is then obtained from
Eq.(\ref{E2}) and (\ref{d16}).

For $n=1$, we find from
Eqs.(\ref{epsilon}), (\ref{d18}), (\ref{d19}) and (\ref{d20})
that
\begin{eqnarray}
\Gamma+2&=&\frac{E-m}{E+m}\frac{(Z\alpha)^2}{x_0^2}~,
\label{d21}\\
b_0+\frac{L}{x_0}&=&\frac{2\beta}{x_1}~,
\label{d22}\\
\frac{1}{x_1+x_0}&=&x_0+\frac{L}{x_0}~,
\label{d23}\\
\frac{2\beta}{x_1}-x_1&-&\frac{1}{x_1+x_0}=0~.
\label{d24}
\end{eqnarray}
Eq.(\ref{d22}), (\ref{d23}), and (\ref{d24}) imply
$x_1=b_0-x_0$.
Substituting $x_1$ into Eq.(\ref{d23}), we obtain
\begin{eqnarray}
x_0^2=L\left[\frac{E+m}{2E(Z\alpha)^2}-1\right]^{-1}~.
\label{d27}
\end{eqnarray}
Then from Eq.(\ref{d27}) and Eq.(\ref{d21}), we get
\begin{eqnarray}
\left[4(\Gamma+1)-\frac{\Gamma}{z^2\alpha^2}\right]E^2+4Em+\frac{\Gamma}
{(Z\alpha)^2}m^2=0~.
\label{d28}
\end{eqnarray}
The energy $E$ can be solved from Eq.(\ref{d28}) by the standard
formula, after which the magnetic field is determined from
Eq.(\ref{E2}). Eq.(\ref{d28}) does not resemble the one obtained
from recursion relation in \cite{HoKha}.  However, on multiplying
Eq.(\ref{d28}) by $\Gamma+1$ and making use of the fact that
$(Z\alpha)^2=\Gamma(\Gamma-2\gamma)$, we can show, after some
algebra, that Eq.(\ref{d28}) is equivalent to the corresponding
equation given in \cite{HoKha}.

Finally we consider the case for $n=2$.  We have Eq.(\ref{epsilon}) with
$\epsilon=2$, together with
Eqs.(\ref{d18}), (\ref{d19}) and (\ref{d20}) in the forms
\begin{eqnarray}
\Gamma+3&=&\frac{E-m}{E+m}\frac{(Z\alpha)^2}{x_0^2}~,
\label{d30}\\
b_0+\frac{L}{x_0}&=&\frac{2\beta}{x_1}+\frac{2\beta}{x_2}~,
\label{d31}\\
\frac{1}{x_1+x_0}&+&\frac{1}{x_2+x_0}=x_0+\frac{L}{x_0}~,
\label{d32}\\
\frac{2\beta}{x_1}-x_1&-&\frac{1}{x_1+x_0}-\frac{2}{x_2-x_1}=0~,
\label{d33}\\
\frac{2\beta}{x_2}-x_2&-&\frac{1}{x_2+x_0}-\frac{2}{x_1-x_2}=0~.
\label{d34}
\end{eqnarray}
From these equations we find $x_1+x_2=b_0-x_0$ and $x_1x_2=2\beta
x_0(b_0-x_0)/(b_0x_0+L)$. Putting these expressions into
Eq.(\ref{d32}) and using the fact that $\Gamma=2\beta+L-1$, we
arrive at
\begin{eqnarray}
\left(b_0^2-2\beta\right) x_0^2 + b_0\Gamma x_0
+\left[b_0^2\left(L-1\right)-L\left(2\beta+1\right)\right]+
\frac{b_0 \Gamma L}{x_0}=0 ~.
\label{38}
\end{eqnarray}
Now multiplying Eq.(\ref{38}) by $\Gamma$, using $\Gamma L=(Z\alpha)^2$, and
expressing $b_0$, $l_B$, and $1/x_0^2$ in terms of $E$, we get finally
\begin{eqnarray}
\left\{4(2\Gamma+3)-\frac{1}{(Z\alpha)^2}\left[6\Gamma+2(\gamma+1)
+\frac{(2\gamma+1)\Gamma}{\Gamma+3}\right]\right\}~E^3\nonumber\\
+\left\{12-\frac{1}{(Z\alpha)^2}\left[2(\gamma+1)-\frac{(2\gamma+1)
\Gamma}{\Gamma+3}\right]\right\}~E^2m\nonumber\\
+ \frac{1}{(Z\alpha)^2}\left[6\Gamma+2(\gamma+1)
+\frac{(2\gamma+1)\Gamma}{\Gamma+3}\right]~Em^2~~\nonumber\\
+\frac{1}{(Z\alpha)^2}\left[2(\gamma+1)-\frac{(2\gamma+1)
\Gamma}
{\Gamma+3}\right]~m^3=0~.
\label{d39}
\end{eqnarray}
Again, this equation does not look the same as that obtained from the
recursion relations.  But we can show they are in fact equivalent
as they differ only by a multiplicative factor
$(\Gamma+1)(\Gamma+2)$.

\vskip 0.5 truecm

{\bf 5.}  We now demonstrate that the QES equation (\ref{d6})
cannot be represented as bilinear combination of the generators of
the $sl_2$ algebra.  The question of whether there exists
non-$sl_2$-based one-dimensional QESM was first posed in
\cite{Tur2} in which all $sl_2$-based QESM are classified. The
first example of such a kind was given in \cite{JKK}, which
presents a potential arising in the context of the stability
analysis around the kink solution for $\phi^4$-type field theory
in $1+1$ dimensions.

To show that Eq.(\ref{d6}) is also not generated by the $sl_2$
algebra, let us rewrite it as
\begin{eqnarray}
\left\{-\left(x^2+x_0x\right)\frac{d^2}{dx^2}+\left[x^3+x_0x^2+\left(1-2\beta
\right)x-2\beta x_0\right]\frac{d}{dx}\right. \nonumber\\
\left.-\epsilon x^2 +\left(c-b-\epsilon
x_0\right)x-bx_0\right\}Q(x)=0~.
\label{d6-2}
\end{eqnarray}
Turbiner \cite{Tur2} has shown that all $sl_2$-based second order
QES differential equations can be cast into the form
\begin{eqnarray}
-P_4(x)\frac{d^2 Q}{dx^2}+
P_3(x)\frac{dQ}{dx}+\left(P_2(x)-\lambda\right)Q=0~, \label{qesm}
\end{eqnarray}
where
\begin{eqnarray}
P_4(x)&=&a_{++}x^4+a_{+0}x^3+\left(a_{+-}+a_{00}\right)x^2+a_{0-}x+a_{--}~,
\nonumber\\
P_3(x)&=&2\left(2j-1\right)a_{++}x^3+\left[\left(3j-1\right)a_{+0}+b_+\right]
 x^2\nonumber\\
&+& \left[2j\left(a_{+-}+a_{00}\right)+a_{00}+b_0\right]x+ja_{0-}+b_-~,
\nonumber\\
P_2(x)&=&2j\left(2j-1\right)a_{++}x^2 +
2j\left(ja_{+0}+b_+\right)x+a_{00}j^2+b_0j~.
\label{P}
\end{eqnarray}
Here $a_{kl}$'s and $b_k$'s ($k,l=+,0,-$) are constants, and $j$
is a non-negative integer or half-integer. Eq.(\ref{qesm})
corresponds to the eigenvalue problem
\begin{eqnarray}
HQ=\lambda Q~, \quad\quad H=-\sum_{k,l=+,0,-\atop k\geq l}
a_{kl}J^kJ^l +\sum_{k=+,0,-} b_k J^k~,
\end{eqnarray}
which has a polynomial solution of power $2j$ in $x$. Here $J^k$'s
are the generators of $sl_2$:
\begin{eqnarray}
J^+=x^2\frac{d}{dx}-2jx~,\quad J^0=x\frac{d}{dx}-j~,\quad J^-=\frac{d}{dx}~.
\end{eqnarray}
Comparing Eqs.(\ref{d6-2}) and (\ref{qesm}) we find that the two equations
are inconsistent with each other.  For instance, the coefficient of $x^4$ in
$P_4$ requires $a_{++}=0$, whereas the coefficient of $x^3$ in $P_3$ implies
$2(2j-1)a_{++}=1$, which gives a non-vanishing $a_{++}$ for positive integral
and half-integral values of $j$.  This shows that Eq.(\ref{d6}) is not
$sl_2$-based.

\vskip 0.5 truecm

{\bf 6.}  In conclusion,  we have given an algebraic solution to
the planar Dirac equation of an electron in the Coulomb and
magnetic fields.  The relevant Bethe ansatz equations are
presented.  Unlike the corresponding equations in the
Schr\"odinger and the Klein-Gordon case discussed in \cite{ChHo},
the unknown variables in this set of Bethe ansatz equations
include not only the roots of the polynomial assumed, but also a
parameter from the QES differential operator.  Equivalence between
this approach and that by the recursion relations is demonstrated.
Finally, we show that the QES equation for this problem does not
belong to any of the classes based on the $sl_2$ algebra.

\vskip 3cm
\centerline{\bf Acknowledgment}

This work was supported in part by the Republic of China through
Grant No. NSC 89-2112-M-032-020.

\end{document}